\documentclass[10pt]{article}
\usepackage{geometry}                
\usepackage{graphicx}
\usepackage{amssymb}
\usepackage{epstopdf}
\usepackage{lineno}
\usepackage{authblk}
\usepackage{multicol}
\usepackage{amsmath}
\usepackage{mathrsfs}
\usepackage{graphicx}
\usepackage{cite}
\usepackage{hyperref}
\newcommand{\sqeqref}[1]{Eq. \ref{#1}}
\newcommand{\figref}[1]{Fig. \ref{#1}}

\title{Exploring the interplay of helicity and global-frame spin density matrix elements}

\author[1]{Gavin Wilks}
\author[2]{Zhenyu Ye}
\affil[1]{\textit{University of Illinois at Chicago, Chicago, Illinois 60607, USA}}
\affil[2]{\textit{Lawerence Berkeley National Laboratory, Berkeley, California 94720, USA}} 
\date{}

\begin{document}
\maketitle
\section*{Abstract}
A significant $\phi$-meson global spin alignment ($\rho_{00}$) signal was measured in Au+Au collisions at $\sqrt{s_{NN}}\le62$ GeV \cite{Nature}. Conventional physics mechanisms
fail to accommodate this $\rho_{00}$ signal, motivating further investigation into its origin. Recent studies from a general framework utilizing the gauge/gravity duality predict that there could be non-zero helicity-frame spin alignment generated by the relative motion of $s\bar{s}$ pairs to the thermal background in heavy-ion collisions, leading to significant global $\rho_{00}$ \cite{HelicityFrame}. In this work, we derive the expected relationships between $\phi$-meson helicity and global-frame spin density matrix elements and confirm these analytical results using Monte Carlo simulations.
The effects on these relationships from finite elliptic flow and various kinematic selections are also examined. Additionally, we extract $\rho_{00}$ simultaneously with the off-diagonal elements from the 2-dimensional $\theta^{*}$ and $\beta$ angular dimensions, as opposed to the standard 1-dimensional extraction of $\rho_{00}$. These angles describe the direction of a daughter kaon's momentum in the $\phi$-meson's rest frame, where $\theta^{*}$ is the polar angle with respect to the chosen spin-quantization axis, and $\beta$ is the azimuthal angle in the perpendicular plane.
This 2-dimensional method allows us to consider possible effects from finite geometric acceptance and detector efficiencies, which could couple to the spin density matrix elements and influence the extraction of these elements within the same frame.
The studies presented in this work explore the relationships between helicity and global-frame spin density matrix elements and possible detector effects on the 2-dimensional extraction of spin density matrix elements.


\section{Introduction}

A large orbital angular momentum is generated in non-central heavy-collisions, leading to vorticity of the Quark Gluon Plasma (QGP) along the orbital angular momentum direction. Through spin-orbit couplings \cite{Original}, a particle in the QGP medium can be polarized along this direction, known as global polarization. According to the flavor-spin wave function, the total polarization of a $\Lambda(\bar{\Lambda})$ hyperon is carried solely by the strange quark $s(\bar{s})$; therefore, $s(\bar{s})$ polarization can be measured directly from $\Lambda(\bar{\Lambda})$ polarization \cite{Close}. The STAR Collaboration recently measured a significant $\Lambda(\bar{\Lambda})$ hyperon global polarization signal, providing an indirect probe of QGP vorticity and viscosity \cite{NaturePolarization,LambdaCentrality2}, while also furthering our understanding of particle production mechanisms during hadronization \cite{Original}.

Following these global polarization measurement, vector meson global spin alignment gained much interest as an additional probe of QGP fluid properties and particle production mechanisms within the QGP medium. The $\phi$-meson $(s\bar{s})$ global spin alignment was measured by the STAR Collaboration and significant signal was reported for Au+Au collisions at $\sqrt{s_{NN}}\le62$ GeV \cite{Nature}. Conventional mechanisms fail to accommodate the observed signal, including contributions from $s(\bar{s})$ global polarization in the quark coalescence picture \cite{Magnetic,Helicity,Axial,Vorticity}. Several models were developed in attempts to understand this signal. One such model proposes the existence of a $\phi$-meson strong force field where field fluctuations induce global spin alignment \cite{Sheng1,Sheng2,Sheng3,Sheng4,Sheng5}. Another more recent model utilizes the gauge/gravity duality and suggests global spin alignment originates from a non-zero helicity spin alignment induced by the relative motion of $s\bar{s}$ pairs to the thermal background in heavy-ion collisions \cite{HelicityFrame}. Using this model, predictions were generated for the $\phi$-meson global spin alignment originating from the helicity spin alignment, where the signal is comparable to current measurements. 

In this paper, 
we analytically solve for relations of global and helicity frame spin density matrix elements between and verify with Monte Carlo simulations for various kinematic selections. Understanding possible contributions to global off-diagonal elements is important as they may pose as a physics background when measuring chiral magnetic effect (CME) observables involving vector meson decay products \cite{OffDiagonalOrig,OffDiagonal}. We explore how these relations are modified by the elliptic flow in heavy-ion collisions. We also study the effects of finite geometric acceptance and detector efficiency when measuring these observables and practical methods of correcting for these effects. 

\section{Relationships between global and helicity frame matrix elements}
The first step in relating the helicity frame and global frame spin density matrix elements is to understand the relationship between their coordinate systems, such that we can map one onto the other. We start with the lab frame, where we define the axes to coincide with axes from heavy-ion collision experiments. The $z$-axis is the beam axis ($\hat{p}_{1}$), along which the two nucleons collide, while the $x$ and $y$ axes define the plane perpendicular to the beam axis. This is the frame where our momenta are initially defined. 

In the global frame, the kaon is boosted into the rest frame of the parent $\phi$-meson. We define $\theta^{\ast}_{g}$ as the polar angle of the boosted $K^{+}$ momentum ($\vec{p}_{K^{+}}^{\,\prime}$) with respect to the total angular momentum ($\hat{L}$) and $\beta_{g}$ is the azimuthal angle between the positive beam axis ($\hat{p}_{1}$) and the projection of $\vec{p}_{K^{+}}^{\,\prime}$ onto the reaction plane spanned by $\hat{p}_{1}$ and the impact parameter ($\hat{b}$). The azimuthal angle of $\hat{b}$ in the lab frame is referred to as the reaction plane angle, $\Psi$

In the helicity frame, the kaon is boosted into the rest frame of the parent $\phi$-meson. The direction of the quantization axis ($\hat{z}$) in the helicity frame is defined as the sum of the momenta of the two colliding nucleons in the phi-meson rest frame, $\vec{p}_{1}^{\,\prime}$ and $\vec{p}_{2}^{\,\prime}$. This is anti-parallel to the direction of the $\phi$-meson momentum ($\vec{p}_{\phi}$) in the lab frame. The $y$-axis is defined as perpendicular to the production plane, $\hat{y}=(\vec{p}_{1}^{\,\prime}\times\vec{p}_{2}^{\,\prime})/|\vec{p}_{1}^{\,\prime}\times\vec{p}_{2}^{\,\prime}|$, while the $x$-axis completes the right handed coordinate system, $\hat{x}=\hat{y}\times\hat{z}$. Therefore, $\theta^{\ast}_{h}$ is the polar angle with respect to -$\vec{p}_{\phi}$ and $\beta_{h}$ is the azimuthal angle between $\hat{x}$ and the projection of $\vec{p}_{K^{+}}^{\,\prime}$ onto the $x,y$ plane in the helicity frame.

Following \cite{Distribution2D}, the angular distributions of the $K^{+}$ in the global and helicity frames including all spin density matrix element can be written as, 
\begin{align}
    \frac{d^{2}N}{d\cos{\theta^{\ast}}d\beta} = 
    &\frac{3}{8\pi}\bigr[\left(1-\rho_{00}\right)+\left(3\rho_{00}-1\right)\cos^{2}{\theta^{\ast}} \notag \\
    &-\sqrt{2}\text{Re}\left(\rho_{10}-\rho_{0-1}\right)\sin{2\theta^{\ast}}\cos{\beta} \notag \\
    &+\sqrt{2}\text{Im}\left(\rho_{10}-\rho_{0-1}\right)\sin{2\theta^{\ast}}\sin{\beta} \notag \\
    &-2\text{Re}\left(\rho_{1-1}\right)\sin^{2}{\theta^{\ast}}\cos{2\beta} \notag \\
    &+2\text{Im}\left(\rho_{1-1}\right)\sin^{2}{\theta^{\ast}}\sin{2\beta} \bigr],
    \label{eq:2D}
\end{align}
where the helicity and global frame angles can be substituted into this general form. The spin density matrix elements will also differ in each frame and will be distinguished in this paper as $\rho_{\lambda_{1}\lambda_{2}}^{g}$ and $\rho_{\lambda_{1}\lambda_{2}}^{h}$ for the global and helicity frames, respectively, with $\lambda_{1},\lambda_{2}\in\{0,\pm1\}$.

The relationship between the global and helicity frame spin density matrix elements can be derived using the 3-dimensional rotation operator, 
\begin{equation}
    R(\alpha,\beta,\gamma)=e^{-i\gamma J_z}e^{-i\beta J_x}e^{-i\alpha J_y},
    \label{eq:rotations}
\end{equation}
where $J_{x,y,z}$ are the total angular momentum operators in the spin-1 basis, and $\alpha$, $\beta$, and $\gamma$ are the Euler angles for a $y$-$x$-$z$ series of rotations. For our definitions of the global and helicity frame axes, these angles are 
\begin{equation}
    \alpha=\arctan{\left(\frac{p_{z,\phi}}{p_{T,\phi}}\right)},
    \,\, \beta=\frac{\pi}{2}-\left(\phi-\Psi\right), 
    \,\, \gamma=\pi.
    \label{eq:angles}
\end{equation}
The 3-dimension spatial rotations with these given angles align the helicity frame spin quantization axis with $\hat{L}$, which is the spin quantization axis in the global frame. The perpendicular plane is also aligned, such that axes where the $\beta_{h}$ and $\beta_{g}$ angles are defined will exactly coincide.
The rotation operators can be applied to the helicity spin density matrix, providing us with the global spin density matrix,
\begin{equation}
    \rho^{g}=R\left(\alpha,\beta,\gamma\right)\rho^{h}R\left(\alpha,\beta,\gamma\right)^{\dagger}.
    \label{eq:gfromh}
\end{equation}
The resulting matrix is too long to explicitly write here, but we provide the simplest example, where $\rho_{00}^{g}$ can be written as:
\begin{align}
    \rho_{00}^{g}=\frac{1}{2|\vec{p}_{\phi}|^2}&\biggr[|\vec{p}_{\phi}|^{2}\left(1+2\text{Re}\left(\rho_{1-1}^{h}\right)\right)\cos^{2}{\left(\phi-\Psi\right)} +2\sqrt{2}p_{T,\phi}p_{z,\phi}\text{Re}\left(\rho_{10}^{h}-\rho_{0-1}^{h}\right) \notag \\
    &+2p_{T,\phi}^{2}\rho_{00}^{h}
    -p_{z,\phi}^{2}\left(-1+2\text{Re}\left(\rho_{1-1}^{h}\right)+\rho_{00}^{h}\right)\sin^2{\left(\phi-\Psi\right)} \notag \\
    &+|\vec{p}_{\phi}|\left(\sqrt{2}p_{T,\phi}\text{Im}\left(\rho_{10}^{h}-\rho_{0-1}^{h}\right)-2p_{z,\phi}\text{Im}\left(\rho_{1-1}^{h}\right)\right)\sin{\bigr[2\left(\phi-\Psi\right)\bigr]}
    \biggr].
    \label{eq:rhogfromh}
\end{align}
In the next section, we explore the relationships predicted from \sqeqref{eq:gfromh} using Monte Carlo simulations of $\phi$-meson decay.

\section{Monte Carlo Studies}

\subsection{\texorpdfstring{$\phi-\Psi$}{} Dependence}
To test the expected relationships between global and helicity frame spin density matrix elements, we begin by generating $\phi$-mesons with $p_{T,\phi}=2$ GeV/c and $y_{\phi}=1$ at twenty evenly spaced discrete value $\phi-\Psi$ from 0 to $\pi$. We decay $\phi$-meson into $K^{+}K^{-}$ pairs through Pythia and calculate the helicity frame angles of the $K^{+}$ daughter in the $\phi$-meson rest frame. Using \sqeqref{eq:2D}, we simulate effect of the helicity frame spin density matrix elements on the $K^{+}$ daughter distributions by applying $\cos{\theta^{\ast}_{h}}$ and $\beta_{h}$ dependent weights. We then fill the 2-dimensional $\cos{\theta^{\ast}_{g}}$, $\beta_{g}$ distributions of the $K^{+}$ daughters using these weights. The global frame spin density matrix elements are then extracted by fitting the $\cos{\theta^{\ast}_{g}}$, $\beta_{g}$ distributions with \sqeqref{eq:2D} multiplied by a free normalization parameter. 

\begin{figure}[htpb]
    \centering
    \includegraphics[width=\textwidth]{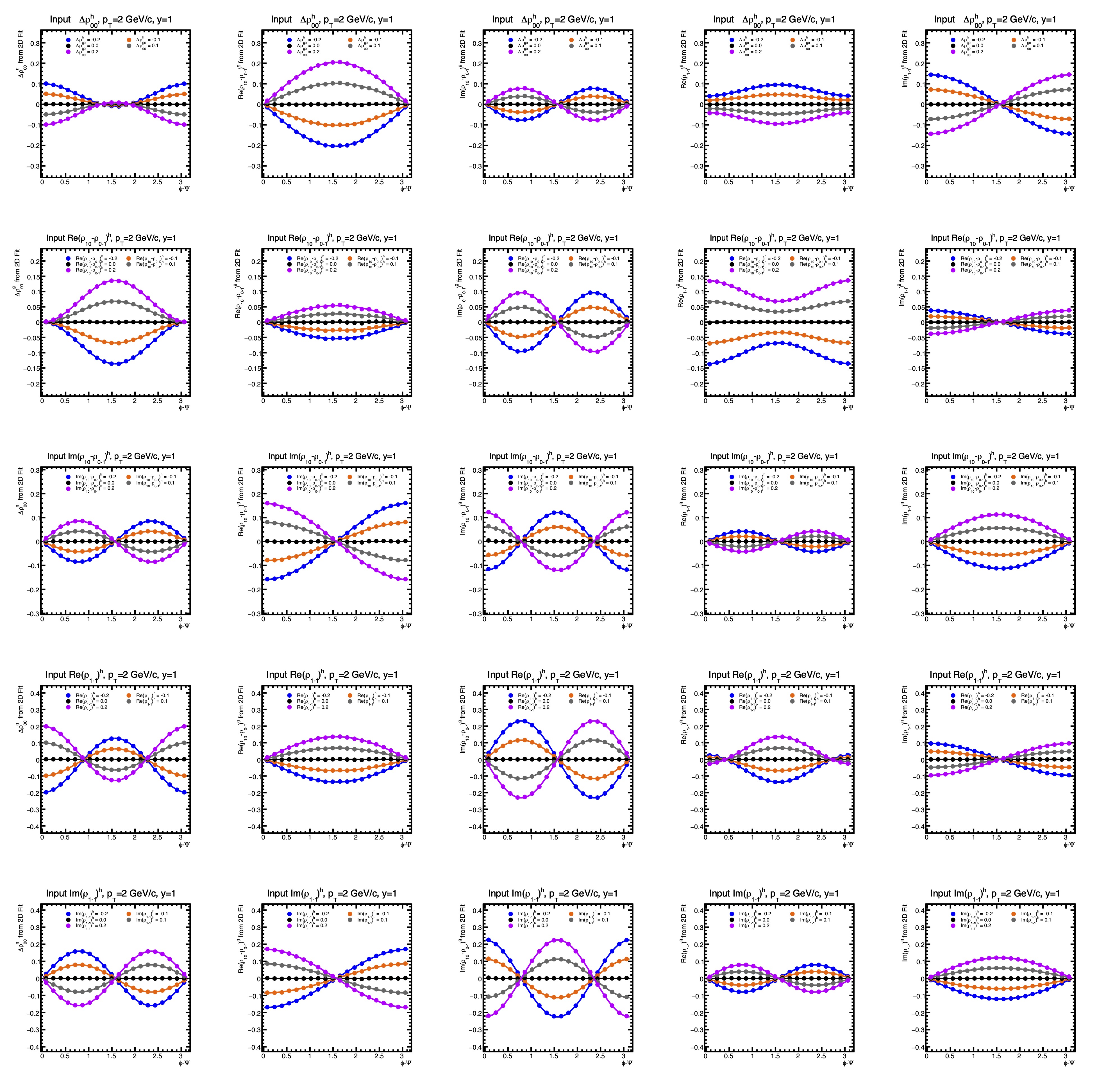}
    \caption{$\phi-\Psi$ dependence for $p_{T,\phi}=2$ GeV/c and $y_{\phi}=1$ of global spin density matrix elements for various inputs of helicity frame spin density matrix elements. Markers represent simulation results, where global parameters are extracted from a fit to $\cos{\theta*_{g}}$ and $\beta_{g}$ using \eqref{eq:2D}. The solid lines are from the analytical relationships from \eqref{eq:gfromh}}.
    \label{fig:phipsi}
\end{figure}

\figref{fig:phipsi} shows the dependence of the extracted $\rho_{\lambda_{1}\lambda_{2}}^{g}$ on the input $\rho_{\lambda_{1}\lambda_{2}}^{h}$ as functions of $\phi-\Psi$. Note that we define $\Delta\rho_{00}=\rho_{00}-\frac{1}{3}$ to quantify deviation of $\rho_{00}$ from its default value of $\frac{1}{3}$. The lines shown through the points are directly calculated from \sqeqref{eq:gfromh}. We observe that our simulations match the expected $\phi-\Psi$ dependence from our calculations. In the case 
of these $\phi$-meson input kinematics, every $\rho^{h}\left(\phi-\Psi\right)$ element affects every $\rho^{g}\left(\phi-\Psi\right)$ element.

\begin{figure}[htpb]
    \centering
    \includegraphics[width=\textwidth]{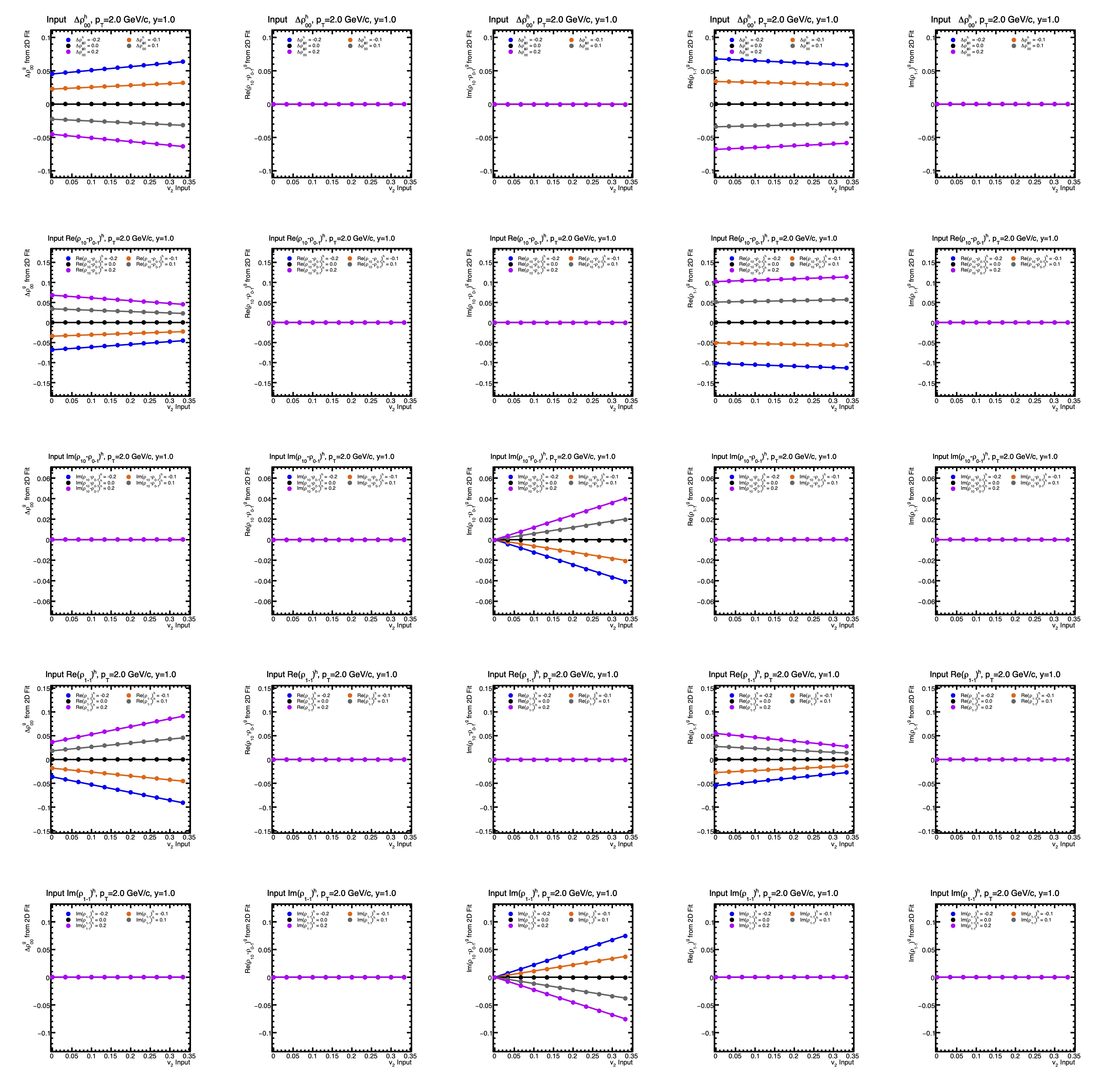}
    \caption{$v_{2}$ dependence for $p_{T,\phi}=2$ GeV/c and $y_{\phi}=1$ of global spin density matrix elements integrated over $\phi-\Psi$ for various inputs of helicity frame spin density matrix elements. Markers represent simulation results, where global parameters are extracted from a fit to $\cos{\theta*_{g}}$ and $\beta_{g}$ using \eqref{eq:2D}. The solid lines are from the analytical relationships from \eqref{eq:gfromh}}
    \label{fig:v2}
\end{figure}

\begin{figure}[htpb]
    \centering
    \includegraphics[width=\textwidth]{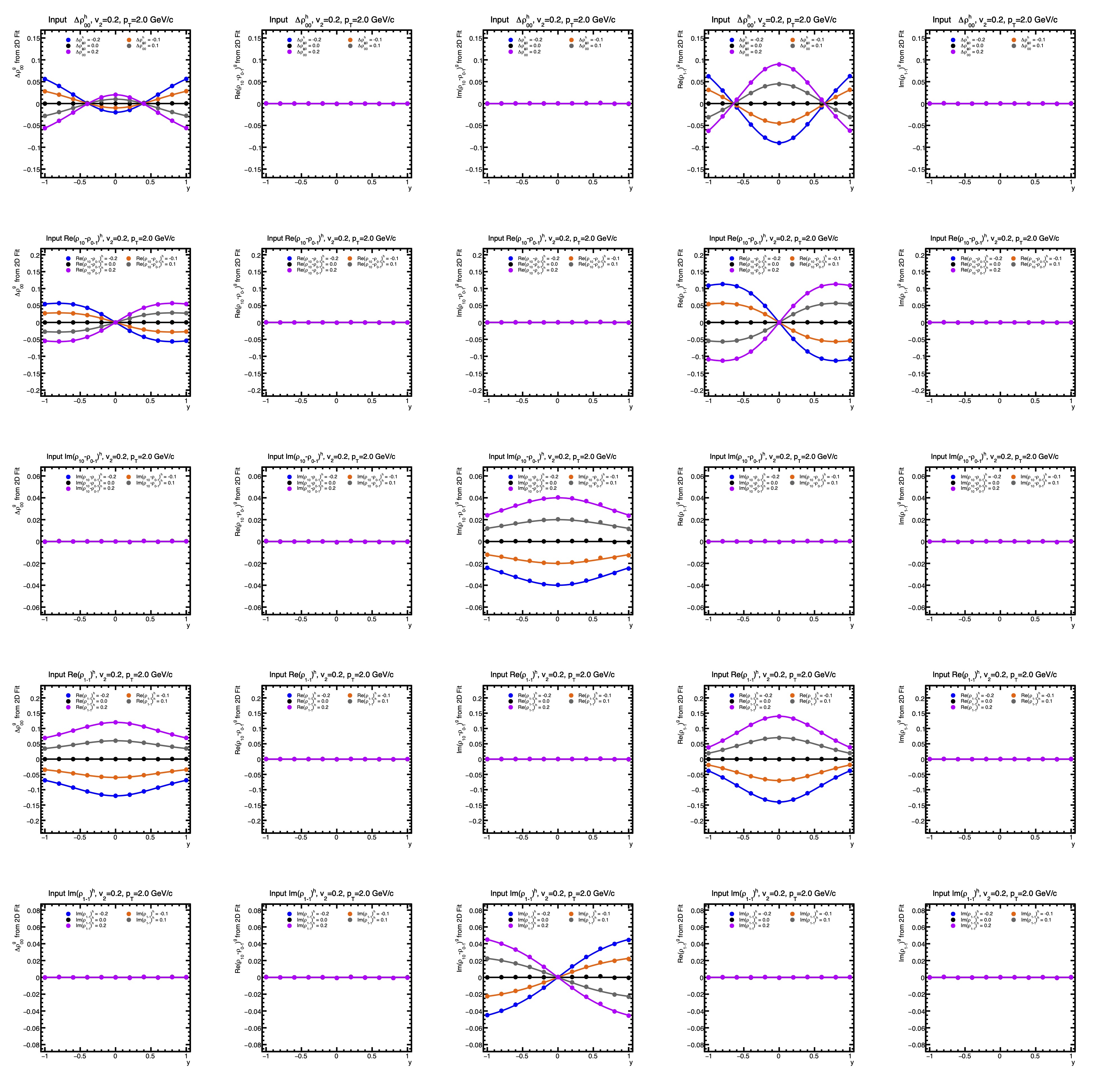}
    \caption{$y_{\phi}$ dependence for $p_{T,\phi}=2$ GeV/c and $v_{2}=0.2$ of global spin density matrix elements integrated over $\phi-\Psi$ for various inputs of helicity frame spin density matrix elements. Markers represent simulation results, where global parameters are extracted from a fit to $\cos{\theta*_{g}}$ and $\beta_{g}$ using \eqref{eq:2D}. The solid lines are from the analytical relationships from \eqref{eq:gfromh}}
    \label{fig:rapidity}
\end{figure}

\begin{figure}[htpb]
    \centering
    \includegraphics[width=\textwidth]{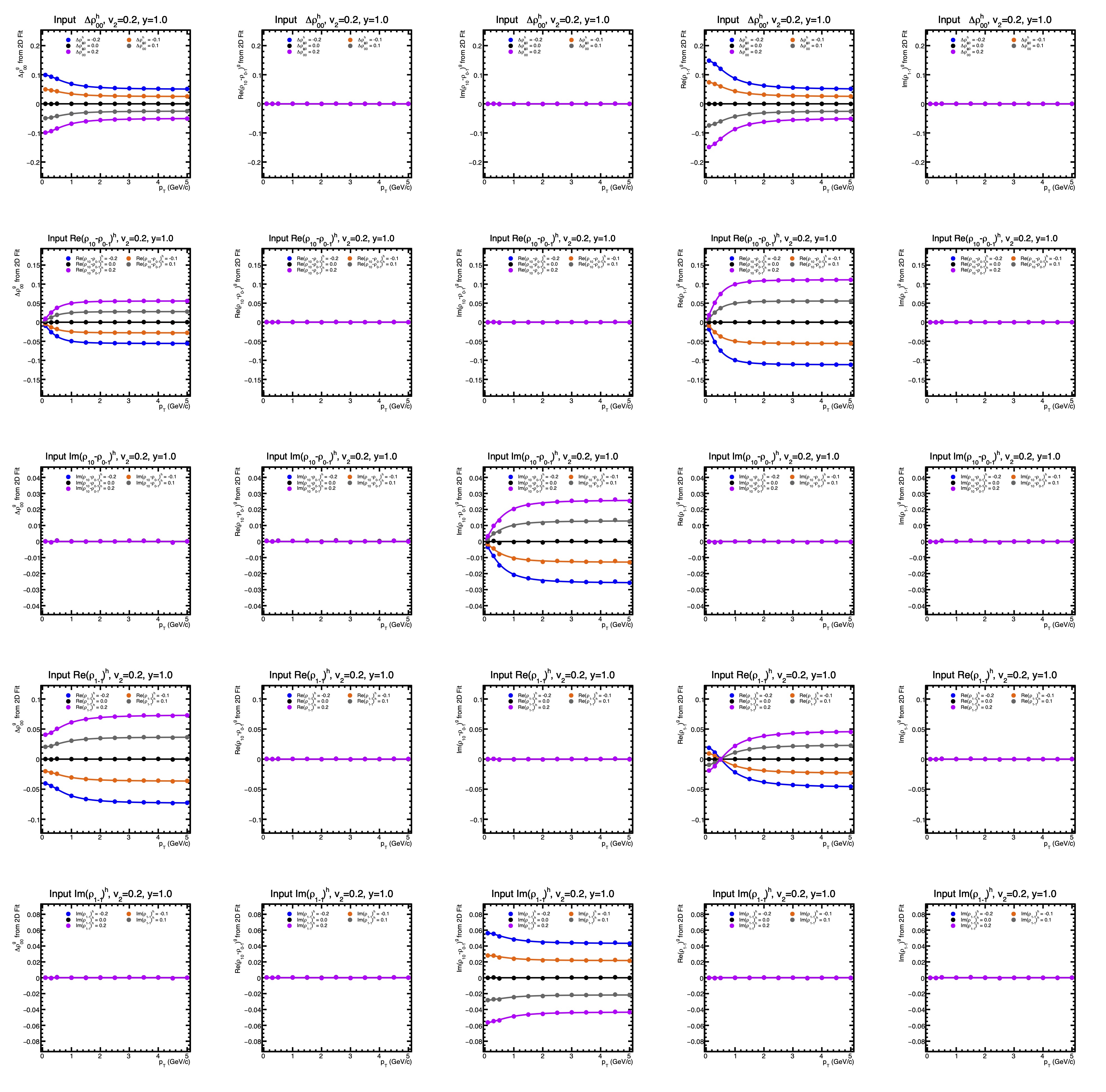}
    \caption{$p_{T,\phi}$ dependence for $y_{\phi}=1$ and $v_{2}=0.2$ of global spin density matrix elements integrated over $\phi-\Psi$ for various inputs of helicity frame spin density matrix elements. Markers represent simulation results, where global parameters are extracted from a fit to $\cos{\theta*_{g}}$ and $\beta_{g}$ using \eqref{eq:2D}. The solid lines are from the analytical relationships from \eqref{eq:gfromh}}
    \label{fig:pt}
\end{figure}

\subsection{Effects from Elliptic Flow}

Elliptic flow describes the degree of anisotropy in the angular distribution of particle momenta produced in heavy-ion collisions. Using Fourier decomposition the azimuthal distribution for any given particle can be expressed as,
\begin{equation}
    \frac{dN}{d\phi}=\frac{1}{2\pi}\left(1+\sum_{n=0}^{\infty} 2v_{n}\cos{\bigr[n\left(\phi-\Psi\right)]}\right).
    \label{eq:flow}
\end{equation}
Elliptic flow is quantified by the $n=2$ coefficient in the sum, $v_{2}$ \cite{flow1,flow2,flow3,flow4,flow5}. 
Previous measurements of vector meson global spin alignment integrate over the full $\phi-\Psi$ dependence; therefore, predictions of the global spin alignment from helicity frame parameters requires the same. We can integrate \sqeqref{eq:gfromh} over $\phi-\Psi$ weighted by \sqeqref{eq:flow} to find the coupling effect of a finite $v_{2}$ with the helicity frame parameters. We denote the average over $\phi-\Psi$ as $\langle\rangle$. As an example, $\langle\rho_{00}^{g}\rangle$ can be found by averaging \sqeqref{eq:rhogfromh}, arriving at
\begin{align}
    \langle\rho_{00}^{g}\rangle=
    \frac{1}{8}&\biggr[3+2\text{Re}\left(\rho_{1-1}^{h}\right)-\rho_{00}^{h}+v_{2}
    +6\text{Re}\left(\rho_{1-1}^{h}\right)v_{2}-3\rho_{00}^{h}v_{2} \notag \\
    &-\left(-1+2\text{Re}\left(\rho_{1-1}^{h}\right)+3\rho_{00}^{h}\right)\left(-1+v_{2}\right)\left(2\frac{p_{T,\phi}^{2}}{|\vec{p}_{\phi}|^{2}}-1\right) \notag \\
    &-4\sqrt{2}\frac{p_{z,\phi}p_{T,\phi}}{|\vec{p}_{\phi}|^{2}}\text{Re}\left(\rho_{10}^{h}-\rho_{0-1}^{h}\right)\left(-1+v_{2}\right)
    \biggr].
\end{align}
We simulate $\phi$-mesons again at $p_{T,\phi}=2$ GeV/c and $y_{\phi}=1$, but we now use a continuous $\phi-\Psi$ distribution from $-\pi$ to $\pi$. We weight this angular distribution according to \sqeqref{eq:flow} with various  $v_{2}$ inputs. Again, we weight the $\cos{\theta^{\ast}_{g}}$, $\beta_{g}$ distributions with $\cos{\theta^{\ast}_{h}}$ and $\beta_{h}$ dependent weights to simulate the helicity frame spin density matrix elements. 

In \figref{fig:v2}, we present the $v_{2}$ dependence of the $\langle\rho_{\lambda_{1}\lambda_{2}}^{g}\rangle$ with variable $\rho_{\lambda_{1}\lambda_{2}}^{h}$. We find that the integration over $\phi-\Psi$ effectively removes the contributions from several $\rho_{\lambda_{1}\lambda_{2}}^{h}$ to $\langle\rho_{\lambda_{1}\lambda_{2}}^{g}\rangle$. 
We observe that $\langle\rho_{0,0}^{g}\rangle$ and $\langle\text{Re}\left(\rho_{1-1}^{g}\right)\rangle$ depend on all real components of $\rho^{h}$, while $\langle\text{Re}\left(\rho_{10}^{g}-\rho_{0-1}^{g}\right)\rangle$ has no contributions from any component of $\rho^{h}$. Additionally, $\langle\text{Im}\left(\rho_{10}^{g}-\rho_{0-1}^{g}\right)\rangle$ is only dependent on the imaginary components of $\rho^{h}$, and $\langle\text{Im}\left(\rho_{1-1}^{g}\right)\rangle$ is not affected by any $\rho^{h}$ elements.
All observed dependencies are modified by the input $v_{2}$. Our simulations for all input $\rho_{\lambda_{1}\lambda_{2}}^{h}$ and $v_{2}$ values are consistent with our analytical calculations, represented as solid lines.

\subsection{Rapidity and \texorpdfstring{$p_{T}$}{} Dependence}
Due to the $y$-axis rotation in \sqeqref{eq:gfromh}, which depends on $p_{z,\phi}$ and $p_{T,\phi}$ as seen in \sqeqref{eq:angles}, it is expected that our selections of the $\phi$-meson $p_{T,\phi}$ and $y_{\phi}$ will influence the effect of $\rho_{\lambda_{1}\lambda_{2}}^{h}$ on $\langle\rho_{\lambda_{1}\lambda_{2}}^{g}\rangle$. We study these kinematic dependencies using our simulation of $\phi$-meson decay using Pythia described in the previous sections.

In \figref{fig:rapidity}, we show the $y_{\phi}$ dependence of $\langle\rho_{\lambda_{1}\lambda_{2}}^{g}\rangle$ with fixed value of $p_{T,\phi}=2$ GeV/c and $v_{2}=0.2$ for various input values of $\rho_{\lambda_{1}\lambda_{2}}^{h}$ in our simulation. We observe that $y_{\phi}$ affects all non-zero contributions of $\langle\rho_{\lambda_{1}\lambda_{2}}^{g}\rangle$ to $\rho_{\lambda_{1}\lambda_{2}}^{h}$.
Due to the rapidity odd dependence of 
$\langle\rho_{0,0}^{g}\rangle\left(\text{Re}\left(\rho_{10}^{h}-\rho_{0-1}^{h}\right)\right)$, $\langle\text{Im}\left(\rho_{1-1}^{g}\right)\rangle\left(\text{Re}\left(\rho_{10}^{h}-\rho_{0-1}^{h}\right)\right)$, 
and $\langle\text{Im}\left(\rho_{1-1}^{g}\right)\rangle\left(\text{Im}\left(\rho_{1-1}^{h}\right)\right)$, these contributions will integrate to 0 in experimental measurements with symmetric particle selection around $y=0$. The simulation results are consistent with our analytical calculations.

\figref{fig:pt} shows the simulated $p_{T,\phi}$ dependence of our $\phi-\Psi$ integrated global frame spin density matrix elements for various input $\rho_{\lambda_{1}\lambda_{2}}^{h}$. We fix $y_{\phi}=1$ and $v_{2}=0.2$ and find that there is significant $p_{T,\phi}$ dependence of $\langle\rho_{\lambda_{1}\lambda_{2}}^{g}\rangle$ for $p_{T,\phi}\leq2$ GeV/c and then the distributions flatten out towards higher $p_{T,\phi}$. All simulation results are consistent with analytical calculations.





\section{Summary}

In this work, we analytically solve for the expected contributions for vector meson $\rho^{h}$ elements to all $\rho_{g}$ elements and confirm our relations with Monte Carlo simulations. We demonstrate that every element of $\rho^{h}$ will affect every element in $\rho^{g}$, but integerating over the $\phi-\Psi$ dependence, which is typical in experiment, removes much of the dependencies. Namely, it removes contributions from $\rho_{h}$ elements to $\langle\text{Im}\left(\rho_{1-1}^{g}\right)\rangle$ and $\langle\text{Re}\left(\rho_{10}^{g}-\rho_{0-1}^{g}\right)\rangle$, while removing contributions from real and imaginary $\rho^{h}$ elements to imaginary and real $\langle\rho^{g}\rangle$ elements, respectively. We also observe $v_{2}$, $y$, and $p_{T}$ dependence of $\langle\rho^{g}\rangle$ elements for various non-zero values of $\rho^{h}$ elements. 



\end{document}